\documentclass[preprint,aps,prd,nofootinbib,superscriptaddress,tightenlines]{revtex4}
\usepackage{bm}

\usepackage{amsmath}
\usepackage{float}
\usepackage{dcolumn}
\usepackage{amssymb}
\usepackage{graphics}
\usepackage{amsmath, amsfonts, amsthm, amssymb, graphicx,comment}
\usepackage{etex}
\usepackage{hyperref,subfigure,color}

\newcommand{\bea}{\begin{eqnarray}}
\newcommand{\eea}{\end{eqnarray}}

\allowdisplaybreaks[3]

\def \bal#1\eal  {\begin{align} #1 \end{align}}
\newcommand{\be} {\begin{equation}}
\newcommand{\ee} {\end{equation}}
\newcommand{\bpm}{\begin{pmatrix}}
\newcommand{\epm}{\end{pmatrix}}
\newcommand{\nn} {\nonumber\\}

\newcommand{\ud} {\mathrm{d}}

\newcommand{\pd} {\partial}

\newcommand{\mc} {\mathcal}
\newcommand{\tld}{\tilde}

\newcommand{\ai}{{\alpha}}
\newcommand{\bi}{{\beta}}

\newcommand{\ri}{{\rho}}
\newcommand{\si}{{\sigma}}

\begin{document}

\title{Hairy black holes in scalar extended massive gravity}

\author{Andrew J.~Tolley}
\email[]{andrew.j.tolley@case.edu}
\affiliation{Department of Physics, Case Western Reserve University, 10900 Euclid Ave, Cleveland, OH 44106, USA}

\author{De-Jun Wu}
\email[]{wudejun10@mails.ucas.ac.cn}
\affiliation{School of Physics, University of Chinese Academy of Sciences, Beijing 100049, China}

\author{Shuang-Yong Zhou}
\email[]{sxz353@case.edu}
\affiliation{Department of Physics, Case Western Reserve University, 10900 Euclid Ave, Cleveland, OH 44106, USA}

\begin{abstract}

We construct static, spherically symmetric black hole solutions in scalar extended ghost-free massive gravity and show the existence of hairy black holes in this class of extension. While the existence seems to be a generic feature, we focus on the simplest models of this extension and find that asymptotically flat hairy black holes can exist without fine-tuning the theory parameters, unlike the bi-gravity extension, where asymptotical flatness requires fine-tuning in the parameter space. Like the bi-gravity extension, we are unable to obtain asymptotically dS regular black holes in the simplest models considered, but it is possible to obtain asymptotically AdS black holes.

\end{abstract}

\maketitle


\section{Introduction}

The idea of giving the graviton a mass has a long history (see \cite{deRham:2014zqa,Hinterbichler:2011tt} for recent reviews). Apart from theoretical curiosity, massive gravity with a Hubble scale graviton mass may be accountable for the accelerated cosmic expansion observed recently \cite{Riess:1998cb,Perlmutter:1998np}. The unique Poincare-invariant free spin-2 theory, Fierz-Pauli theory, has been discovered long ago and known to be free of ghost instabilities \cite{Fierz:1939ix}. However, the Boulware-Deser ghost \cite{Boulware:1973my} generically arises when adding interactions to promote Fierz-Pauli theory to Lorentz-invariant nonlinear massive gravity. Recently, a two-parameter nonlinear extension to Fierz-Pauli theory has been found \cite{deRham:2010ik, deRham:2010kj} that is free of the Boulware-Deser ghost \cite{deRham:2010kj,Hassan:2011hr,Hassan:2011ea}.

Black hole solutions, particularly static, asymptotically flat black holes, are important both theoretically and phenomenologically to understand a gravitational theory. Black hole solutions in ghost-free massive gravity and bi-gravity have been investigated by various authors \cite{Nieuwenhuizen:2011sq,Koyama:2011xz,Koyama:2011yg,Gruzinov:2011mm,Comelli:2011wq,Berezhiani:2011mt,Volkov:2012wp,Baccetti:2012ge,Cai:2012db,Berezhiani:2013dw,Mirbabayi:2013sva,Volkov:2013roa,Tasinato:2013rza,Babichev:2013una,Brito:2013wya,Berezhiani:2013dca,Brito:2013xaa,Kodama:2013rea,Addazi:2014mga,Babichev:2014oua,Babichev:2014fka,Babichev:2014tfa,Volkov:2014ooa,Babichev:2015xha,Kobayashi:2015yda,Enander:2015}. Since diffeomorphism invariance is explicitly broken in massive gravity, some would-be coordinate singularities in General Relativity (GR) now become physical ones. To see this, it is convenient to restore diffeomorphism invariance by invoking some Stueckelberg scalars $\phi^a$, where $a$ labels different scalars, and observe that there are extra diffeomorphism scalars that have to be kept finite. Since (the components of) $I^{ab}=g^{\mu\nu}\partial _{\mu}\phi^{a}\partial_{\nu} \phi^{b}$ are clearly diffeomorphism scalars in the Stueckelberged version of massive gravity, the usual analyticity requirements for a regular solution also apply to $I^{ab}$. Now, if one chooses unitariy gauge $\phi^a=x^a$,  $I^{ab}$ are nothing but the components of the inverse metric $g^{ab}$. This means that for a solution to be non-singular in massive gravity all the components of $g^{\mu\nu}$ (and thus $g_{\mu\nu}$) have to be non-singular \cite{Berezhiani:2011mt}. For a static, spherically symmetric black hole with spherical coordinates, since its event horizon is necessarily a Killing horizon of $\pd_t$, $g_{tt}=g_{\mu\nu}(\pd_t)^\mu (\pd_t)^\nu$ has to vanish at the horizon. Thus, for such a case, the non-diagonal component $\ud t \ud r$ of the metric should be non-zero to have a regular black hole \cite{Berezhiani:2011mt,Deffayet:2011rh}. This result can be generalized to stationary black holes: since reference metric $\eta_{ab}$ does not have a non-planar Killing horizon, for $g_{\mu\nu}$ to be a regular black hole, the requirement is that the (virtual) bifurcation surface of $g_{\mu\nu}$ can not lie in the interior of the $\eta_{ab}$ spacetime patch (when, e.g., unitary gauge $\phi^a=x^a$ is chosen to identify the spacetime and the internal space) \cite{Deffayet:2011rh}.

Some exact black hole solutions in the literature, which have a vanishing $\ud t \ud r$ term, are singular in the sense discussed above. Nevertheless, solutions free of this kind of singularity have also been found \cite{Koyama:2011xz,Koyama:2011yg,Berezhiani:2011mt,Kodama:2013rea}. There exists a Birkhoff-like theorem that states that all static, spherically symmetric black holes in ghost-free massive gravity in the vaccum are GR-like solutions. That is, they are all some ``coordinate transformed'' forms of the Schwarzschild or Schwarzschild-de Sitter/-Anti-de Sitter geometry \cite{Volkov:2014ooa}. For a given value of the cosmological constant, they would be considered as the same solution in GR, corresponding to different choices of coordinates. But in massive gravity when one performs a coordinate transformation the reference flat metric also changes, which makes it a {\it different} solution. The reason for the absence of non-GR-like solutions is that the staticity ansatz imposes that the effective energy momentum component $X^r_t$ vanish, which implies either the metric is diagonal or the metric is restricted to the case where $X^\mu_\nu$ becomes an effective cosmological constant.

While familiar and simple, these GR-like black holes have been shown to have problems such as singularities, instabilities or strong couplings \cite{Berezhiani:2011mt,Berezhiani:2013dw,Berezhiani:2013dca,Brito:2013wya,Kodama:2013rea}. Therefore, it is desirable to look for well behaved static black hole solutions in generalizations of ghost-free massive gravity. One natural generalization to avoid the staticity condition is to promote the reference metric to a dynamical one \cite{Hassan:2011zd, Hassan:2011ea}. In this extension, when the two metrics are not simultaneously diagonal (i.e., non-bi-diagonal), the black hole solutions are all GR-like \cite{Volkov:2014ooa}. But in bi-gravity the two metrics can be chosen bi-diagonal without encountering the singularities from the extra diffeomorphism scalars ($g^{\mu\nu}f_{\mu\nu}$, etc.), since the two metrics can have the same event horizon in bi-gravity \cite{Deffayet:2011rh}, different from that in massive gravity. The simplest class of solutions in the bi-diagonal case is GR-like solutions where the two metrics are of the same form.
However, these bi-diagonal GR solutions are plagued by the Gregory-Laflamme instability, unless the black hole horizon radius is greater that the graviton Compton length \cite{Babichev:2013una, Brito:2013wya}, which is empirically large, presumably close to the Hubble scale. Interestingly, for the bi-diagonal case, there is a new branch of hairy (non-GR-like) black hole solutions which are asymptotically Anti-de Sitter and have been obtained numerically \cite{Volkov:2012wp}. Asymptotically flat hairy black holes have also been constructed numerically \cite{Brito:2013xaa}, but, unless there is fine-tuning in the theory parameters ($\ai_3$ or $\ai_4$), the horizon radius of this solution is of order of the graviton Compton wavelength, which is phenomenologically unviable. Generically, as extra fields usually inject some gravitating energy in a black hole system, unless there is some kind of protection known as no-hair theorems (see \cite{Bekenstein:1996pn,Herdeiro:2015waa} for a review of no-hair theorems), it might be expected that GR solutions would become unstable in modified gravity theories and a new hairy solution would be the stable solution.

While the bi-gravity extension generalizes ghost-free massive gravity by adding two helicity-2 modes, another class of extended massive gravity models enriches ghost-free massive gravity by simply adding an extra scalar degree of freedom \cite{D'Amico:2012zv,Huang:2012pe,D'Amico:2011jj,Huang:2013mha,DeFelice:2013dua,Andrews:2013ora,Mukohyama:2014rca}. The metric of this class of models has to be non-diagonal to avoid the singularities of the $I^{ab}$ type, but the staticity condition now does not restrict $X^\mu_\nu$ to a cosmological constant anymore. Therefore, one may expect that new branches of hairy black holes be obtained in scalar extended ghost-free massive gravity.

In this paper, we investigate static, spherically symmetric black hole solutions in mass-varying massive gravity \cite{D'Amico:2011jj,Huang:2012pe}. This is a relatively simple class of scalar extended ghost-free massive gravity, but should capture some salient features of black hole solutions in this class of extensions. We will see that, unlike bi-gravity, asymptotically flat hairy solutions can be obtained without fine-tuning the theory parameters for a simple model in this class. To shed some light on the properties of these black holes, we shall also briefly investigate the dependence of the black hole mass and the scalar charge on the theory parameters. In addition, while we have not been able to find regular asymptotically de Sitter hairy black holes (like that in the bi-gravity extension), asymptotically anti-de Sitter black holes have been explicitly constructed.

\section{Model and Setup}

As mentioned in the Introduction, we will focus on mass-varying massive gravity \cite{D'Amico:2011jj,Huang:2012pe}. The model is obtained by simply promoting the graviton mass in ghost-free massive gravity to a function depending on a scalar field, and add the standard kinetic and potential term for this scalar:
\be
\label{theth}
S = \frac{{M_P^2 }}{2} \int {d^4 x} \sqrt { - g} \left[ {R + V\left( \varphi  \right) U(\mc{K}) - \frac{1}{2}g^{\mu\nu} \partial _\mu  \varphi \partial_\nu   \varphi  - W\left( \varphi  \right)} \right]  ,
\ee
where $U(\mc{K})\equiv U_2  + \alpha _3 U_3  + \alpha _4 U_4$, and
\be
U_2  = {\cal K}_{[\mu }^\mu  {\cal K}_{\nu ]}^\nu  , ~~~U_3  = {\cal K}_{[\mu }^\mu  {\cal K}_\nu ^\nu  {\cal K}_{\rho ]}^\rho  , ~~~U_4  = {\cal K}_{[\mu }^\mu  {\cal K}_\nu ^\nu  {\cal K}_\rho ^\rho  {\cal K}_{\sigma ]}^\sigma   ,
\ee
with ${\cal K}^\mu_\nu$ defined as ${\cal K}_\nu ^\mu   = \delta _\nu ^\mu   - \sqrt {g^{\mu \rho }  \eta_{\ri\nu}}$. $\eta_{\ri\nu}$ is the reference Minkowski metric, which explicitly breaks diffeomorphism invariance. If desired, which is not the approach we take to numerical construct solutions in this paper, one can restore diffeomorphism invariance by introducing four stuckelberg fields $\phi^a$ and making the replacement: $\eta_{\ri\nu}\to \pd_\ri\phi^a\pd_\nu\phi^b\eta_{ab}$. $\alpha_3$ and $\alpha_4$ are two free parameters of the model, and $V(\varphi)$ and $W(\varphi)$ are arbitrary functions. As we shall see later, to obtain the hairy black hole solutions, we only need a simple monomial form of $V(\varphi)$ and can simply set $W(\varphi)=0$. Since $V(\varphi)$ regulates the mass of graviton, it has to be positive to avoid tachyonic instabilities. Note that $M_P^2$ is pulled out of the entire integral, so $\varphi$ is dimensionless and $V(\varphi)$ is dimensionally mass squared. We find it convenient to work with the eigenvalues of matrix $\cal K_\nu^\mu$, which are assumed to be $k_1,k_2,k_3$ and $k_4$. Then $U_2$, $U_3$ and $U_4$ can be recast as elementary symmetric polynomials of $k_i$:
\be
\label{defU2}
U_2={\sum\limits_{\nu<\mu} {{k_\nu}{k_\mu}} },~~~
U_3={\sum\limits_{\nu<\mu<\rho} {{k_\nu}{k_\mu}{k_\rho}}},~~~
U_4={k_1}{k_2}{k_3}{k_4}  .
\ee

To derive the Einstein field equation, one can make use of the formula $\pd  [{{\cal K}^n }]/{\pd g^{\mu\nu} } = {n}\left( {{\cal K}^n_{\mu\nu}  - {\cal K}^{n - 1}_{\mu\nu} } \right)/2$, where $ {\cal K}^n_{\mu\nu}=g_{\mu\mu_1}\mc{K}^{\mu_1}_{\mu_2}\mc{K}^{\mu_2}_{\mu_3}\cdots\mc{K}^{\mu_n}_{\nu}$ and $[~]$ denotes the trace of the matrix enclosed. The equations of motion are
\bal
\label{eisteq}
G_{\mu\nu} &=\frac{1}{2}T_{\mu\nu}+V X_{\mu\nu} ,
\\
\label{eompsi}
\partial _\mu \left( {\sqrt {-g } g^{\mu\nu}  \partial_\nu \varphi } \right) &= {\sqrt { -g } } \left( W_\varphi-V_\varphi U\right)   ,
\eal
where $W_\varphi=\pd W/\pd \varphi$, $V_\varphi=\pd V/\pd \varphi$, $T_{\mu\nu}$ is  the energy-momentum tensor of the scalar field
\be
T_{\mu\nu} = \partial_\mu \varphi \partial _\nu \varphi  - g_{\mu\nu}\left(  \frac{1}{2}g^{\ri\si} \partial _\ri  \varphi \partial_\si   \varphi + W \right)  ,
\ee
and the effective energy-momentum tensor coming from the graviton potential is
\bal
 X_{\mu\nu}  &= \frac{1}{2}\bigg[-({\cal K}_{\mu\nu}  - [{\cal K}] g_{\mu\nu})  + \alpha \left( {{\cal K}_{\mu\nu}^2  - [{\cal K}] {\cal K}_{\mu\nu}  + U_2 g_{\mu\nu} } \right)
 \nn
&~~~~~~~~~  - \beta \left( {{\cal K}_{\mu\nu}^3  - [{\cal K}] {\cal K}_{\mu\nu}^2  + U_2 {\cal K}_{\mu\nu}  - U_3 g_{\mu\nu} } \right)\bigg]  .
\eal
Note that we have introduced two new parameters to replace $\alpha_3$ and $\alpha_4$:
\be
\alpha=1+\alpha_3, ~~~~ \beta=\alpha_3+\alpha_4  ,
\ee
which we will use in the rest of the paper.

We are interested in static, spherically symmetric black holes. The most general ansatz satisfying these symmetries is
\begin{align}
\label{metric}
\ud s^2  &=  - a(r)dt^2 + 2b(r)drdt  + e(r)dr^2   + f(r)d\Omega ^2,
\\
\label{metric2}
\ud s^2_\eta &= \eta_{\mu\nu}dx^{\mu}dx^{\nu}  =  - dt^2  + dr^2   +r^2 d\Omega ^2  ,
\\
\label{varphian}
\varphi &= \varphi(r)    .
\end{align}
Although $\ud s^2$ and $\ud s_{\eta}^2$ are both static metrics, the time-like Killing vector $\pd_t$ is $\ud s_{\eta}^2$ orthogonal but not $\ud s^2$ orthogonal to the constant $t$ surfaces. It may be worth emphasizing that when performing a coordinate transformation, it should be applied to both of the metrics simultaneously. A coordinate transformation of metric $\ud s^2$, or $\ud s_{\eta}^2$, {\it alone} leads to a physically inequivalent configuration.

It is straightforward to calculate the nonzero components of $G^\mu_\nu$ and $T^\mu_\nu$ (indices being raised by $g^{\mu\nu}$) for ansatz (\ref{metric})--(\ref{varphian}). To calculate $X^{\mu}_{\nu}$, we note that $k_1,k_2,k_3$ and $k_4$ are
\bal
k_1  &= 1 - \sqrt{ \frac{2}{ {a + e + \sqrt {\left( {a + e} \right)^2  - 4c} }  } }   ,
\\
k_2 &= 1 - \sqrt{ \frac{2}{ {a + e - \sqrt {\left( {a + e} \right)^2  - 4c} }  } }   ,
\\
\label{k3k4eq}
k_3&=k_4=1 - \sqrt {\frac{r^2}{f}},
\eal
where we have defined
\be
\label{cdef}
c\equiv a e+b^2   .
\ee
For $g_{\mu\nu}$ to have the correct Lorentzian signature we need $c>0$ and $f>0$. Also, for $k_1$ and $k_2$ to be real numbers, we impose $a+e>0$. The eigenvectors of $k_1, k_2,k_3$ and $k_4$ can be written respectively as
\be
\bpm 1\\bn\\0\\0 \epm,
\bpm bn\\1\\0\\0 \epm,
\bpm 0\\0\\1\\0 \epm,
\bpm 0\\0\\0\\1 \epm  ,
\ee
with $n=2/(\sqrt{(a-e)^2-4 b^2}+a-e)\neq 0$. Note that $(bn)^2\neq1$, otherwise the first and second eigenvector are collinear, implying a singularity for the solution. The nonzero components of $G^\mu_\nu$, $T^\mu_\nu$ and $X^{\mu}_{\nu}$ (indices being raised by $g^{\mu\nu}$) for ansatz (\ref{metric})--(\ref{varphian}) are given explicitly in Appendix \ref{GTXlist}.

We now consider different components of the equations of motion separately. First, since $R^{r}_{t}=T^{r}_{t}=0$, the ${}^r_t$ component of the Einstein equation is simply $X^{r}_{t} = 0$. Since $c\neq 0$, this equation amounts to
\begin{align}
\label{fenlei}
{b\left( {\beta {{k_3}}^2  + 2\alpha {{k_3}} + 1} \right)=0}  ,
\end{align}
Thus, we have two branches of solutions
\begin{itemize}
\item  $b=0$   ,
\item  $\beta {{k_3}}^2  + 2\alpha {{k_3}} + 1 = 0$  .
\end{itemize}
As discussed in the introduction, the branch $b=0$, where the two metrics can be diagonalized simultaneously, necessarily leads to solutions with singularities at the black hole horizon, thus we shall discard this branch. Therefore, the ${}^r_t$ component of the Einstein equation implies $\beta {{k_3}}^2  + 2\alpha {{k_3}} + 1 = 0$. This condition enforces $k_3$ to be constant in terms of $\ai$ and $\bi$ (or equivalently $\ai_3$ and $\ai_4$):
\be
\label{k3eq}
{{k_3}} = \frac{\pm\sqrt{\alpha ^2-\beta }-\alpha }{\beta }  ,
\ee
which implies that
\be
f=\frac{r^2}{(1-k_3)^2}   .
\ee
and $X_t^t  = X_r^r =  {k_3} (\alpha  {k_3}+2)/2$. For the $k_3$ to real and less than 1 ({\it cf.} Eq.~(\ref{k3k4eq})), the $\ai$-$\bi$ parameter space is constrained to be
\be
\label{abpara}
\alpha ^2  \ge \beta , ~~~ \frac{\pm\sqrt{\alpha ^2-\beta }-\alpha }{\beta }<1  .
\ee

Making use of the results above and after some algebra, we can obtain the following closed system of equations of motion:
\bal
\label{ne1}
2c' &= rc\varphi'^2   ,
\\
\label{ne2}
4r ca' -2ra c'+4ac -4 (1-k_3)^2 c^2 &=  2 r^2 c^2[{k_3} (\alpha  {k_3}+2) V - W] ,
\\
\label{ne3}
ac \varphi ''+\frac{ r^2 (2 c a'-a c')+2r a c }{2 (1-k_3)^2 }\varphi ' &=c^2(W_\varphi-V_\varphi U) ,
\eal
where a prime denotes $\pd/\pd r$, $k_3$ is given by Eq.~(\ref{k3eq}) and $U$ is
\be
U= \frac{(1-\alpha)k_3^2 - 2k_3}{\sqrt{c}}+{k_3} (\alpha  {k_3}+2) .
\ee
Eq.~(\ref{ne1}) is the ${}^t_r$ component of the Einstein equation, Eq.~(\ref{ne2}) is obtained by combining the ${}^t_r$ and ${}^r_r$ component of the Einstein equation, and Eq.~(\ref{ne3}) is the $\varphi$ equation of motion. These equations can readily be written as a 4D dynamical system with variable $(a,c,\varphi,\varphi')$. The metric component $e$ can be obtained by algebraically solving the ${}^\theta_\theta$ component of the Einstein equation once $a$, $c$ and $\varphi$ are obtained by solving the system of ODEs (\ref{ne1})--(\ref{ne3}).

\section{Hairy black hole solutions}
\label{sec:hairysol}

One obvious class of solutions for model (\ref{theth}) are that $\varphi$ is constant and solves $W_\varphi - V_\varphi U=0$ at all $r$. In this case, Eq.~(\ref{ne3}) is already solved, and Eqs.~(\ref{ne1}) and (\ref{ne2}) are then the same as that in ghost-free massive gravity, which in turn are the same as that in GR plus potentially a cosmological constant \cite{Volkov:2012wp}. There are also non-GR-like or hairy solutions, which will be constructed in the following. We will focus on the simplest case where $W(\varphi)=0$ and $V(\varphi)$ is a monomial with mirror symmetry $\varphi \to -\varphi$ (a constant plus a monomial in the case of AdS asymptotics). As we shall see, this simple choice already gives rise to a rich solution space of hairy black holes.

\subsection{Approximate solutions near the horizon}

In this subsection we shall derive the approximate black hole solutions near the horizon $r=r_h$, which will be used to setup the ``initial conditions'' to integrate Eqs.~(\ref{ne1})--(\ref{ne3}) outwards and inwards to obtain the full numerical solutions.

Since $a$, $c$ and $\varphi$ have to be analytic around $r_h$, they can be Taylor expanded as:
\bal
\label{ataylor}
a(r)&=\sum_{n=0} {\frac{1}{{n!}}a^{(n)}(r_h) \left( {r - r_h} \right)^n },
\\
\label{ctaylor}
c(r)&=\sum_{n=0} {\frac{1}{{n!}}c^{(n)}(r_h) \left( {r - r_h} \right)^n },
\\
\label{ftaylor}
\varphi(r) &=\sum_{n=0} {\frac{1}{{n!}}\varphi^{(n)}(r_h) \left( {r - r_h} \right)^n }.
\eal
where ${}^{(n)}$ denotes the $n$-th order derivative. For the black hole ansatz (\ref{metric}), the event horizon should be a Killing horizon of vector $\pd_t$, so we have $a(r_h)=-(g_{\mu\nu} (\pd_t)^\mu(\pd_t)^\nu )_{r_h}=0$. Plugging $a(r_h)=0$ into Eq.~(\ref{ne2}), we get
\be
a'({r_h})  = \left. \frac{4 (1-k_3)^2 c + 2 r^2 c[{k_3} (\alpha  {k_3}+2) V - W]}{4r }\right|_{r_h} .
\ee
Eq.~(\ref{ne3}) can be written as
\be
\label{phi11}
\varphi'' = \frac{ c^2(W_\varphi-V_\varphi U) - \frac{ r^2 (2 c a'-a c')+2r a c }{2 (1-k_3)^2 }\varphi' }{ac} .
\ee
Since at the horizon $a$ has to vanish and $c=ae + b^2$ has to remain finite, the denominator in Eq.~(\ref{phi11}) diverges at the horizon. For $\varphi$ to be analytical, the numerator must go to zero as $r\to r_h$. This gives
\be
\varphi'({r_h})  = \left.  \frac{2 (1-k_3)^2 c^2(W_\varphi-V_\varphi U)}{ r^2 (2 c a'-a c')+2r a c } \right|_{r_h} .
\ee
Then Eq.~(\ref{ne1}) gives
\be
c'({r_h}) = \left. \frac{rc\varphi'^2}2 \right|_{r_h} .
\ee
$c({r_h}$) and $\varphi({r_h})$ are undetermined by the equations of motion near the horizon, and thus are potentially free parameters of the black hole solutions. One can differentiate Eqs.~(\ref{ne1})--(\ref{ne3}) with respect to $r$ repeatedly to obtain the higher orders Taylor coefficients $a^{(n)}(r_h),c^{(n)}(r_h),\varphi^{(n)}(r_h)$, which are all determined once $c({r_h})$ and $\varphi({r_h})$ are specified. The expressions for the second and higher order coefficients are lengthy and cumbersome to be displayed here. In this paper, we shall make use of the first two orders in Eqs.~(\ref{ataylor})--(\ref{ftaylor}) as the ``initial'' input in the numerical integration below, the numerical accuracy of which is already sufficient for our purposes.

\subsection{Full numerical solutions}

As mentioned above, we will, for simplicity, choose $W(\varphi)=0$ and only consider monomials of $V(\varphi)$ with mirror symmetry $\varphi \to -\varphi$ in the following. To present numerical solutions, we also need to choose specific values for the dimensionless model parameters $\alpha$ and $\beta$, which have to satisfy
\be
\label{rck3}
\alpha^2\geq\beta,~~~~\frac{(\pm\sqrt{\alpha ^2-\beta }-\alpha)}{\beta} <1
\ee
to ensure the reality of $k_3$. After choosing $\alpha$ and $\beta$, there is also a branch choice for $k_3$, which will be denoted as the $k_{3+}$ and $k_{3-}$ branches.

The apparent ``free'' parameters at the horizon are: $r_h$, $c({r_h}$) and $\varphi({r_h})$. But we can always choose all the dimensionful quantities in units of $r_h$, so $r_h$ factorizes and cancels out in all the equations of motion, which become dimensionless equations, and we are left with  $c({r_h}$) and $\varphi({r_h})$. As we shall see shortly and in the Appendices, to get a weakly asymptotically flat solution ({\it cf.}~Appendix \ref{sec:nonan}), there is no need to do a shooting procedure for $c({r_h}$) and $\varphi({r_h})$. But, to get a Gaussian fall-off asymptotically flat solution, some shooting procedure is required, which will fix at least one of $c({r_h}$) and $\varphi({r_h})$.

With the ``initial conditions'' set up near the horizon, we can integrate Eqs.~(\ref{ne1})--(\ref{ne3}) both outwards and inwards from $r=r_h(1 \pm \epsilon)$ respectively, where $\epsilon$ is chosen to be $10^{-8}$ for the solutions presented in this paper.
The solutions will be plotted down to $r=r_h/10$, but we have  integrated inwards down to $r= r_h/1000$, without encountering any finite radius singularities.

From Eq.~(\ref{ne2}), we can infer that the asymptotics are determined by the value of $VX^t_t=V k_3 \left(\alpha  k_3+2\right)/2$ at spatial infinity (with $W(\varphi)$ chosen to be zero), which acts as an effective cosmological constant asymptotically. Thus, black holes with flat asymptotics may be obtained if $V(\varphi)|_{r= \infty}= 0$, while anti-de Sitter  or de Sitter asymptotics can potentially be achieved when $V k_3 \left(\alpha  k_3+2\right)$ tends to positive and negative values at infinity respectively. However, numerically, we can find solutions with flat and anti-de Sitter asymptotics, but not with de Sitter asymptotics. This is similar to the bi-gravity extension of ghost-free massive gravity where asymptotically anti-de Sitter, but not de Sitter, solutions have also been found. However, in contrast, the asymptotically flat solutions in the bi-gravity extension requires a fine-tuning of the theory parameters $\ai$ and $\bi$ or a fine-tuning of the graviton mass $m$. For the case of the mass fine-tuning, the black hole horizon radius is of order of the graviton Compton wavelength \cite{Brito:2013xaa}. For the scalar extension to be presented below, asymptotically flat black holes exist for generic theory parameters.

We have found numerical solutions for various monomials, as well as for polynomials and/or with nonzero $W(\varphi)$, but in the following we will focus on solutions for the simple case
\begin{itemize}
\item
Asymptotically flat black holes: $V(\varphi)= m^2 \varphi^4$
\item
Asymptotically AdS black holes: $V(\varphi)= v_0 + m^2 \varphi^4$
\end{itemize}
where $v_0$ and $m$ are constants. $m$ has mass dimension 1 and describes the graviton mass for a background configuration with $\varphi\sim\mc{O}(1)$, so phenomenologically viable values of $m$ may be as small as the current Hubble constant. Thus we will consider black holes $r_h\ll 1/m$. Numerically, it is challenging to let $1/m$ and $r_h$ differ by too many orders of magnitude, so we shall construct solutions where $r_h$ is only up to a couple of orders of magnitude smaller that $1/m$. But we see no obstruction for these solutions to be extrapolated down to even smaller $r_h$.

In the following, we will present black hole solutions with flat and AdS asymptotics separately.

\subsubsection{Asymptotically flat black holes}
\label{sec:afs}

As mentioned above, we will consider the case $V(\varphi)= m^2 \varphi^4$ for asymptotically flat black holes. For this case, the condition $V(\varphi)|_{r= \infty}= 0$ implies that $\varphi|_{r= \infty}= 0$.

A generic choice of $c({r_h})$ and $\varphi({r_h})$ leads to asymptotically flat black holes in the weak sense of asymptotical flatness (There are different definitions of asymptotical flatness; see Appendix \ref{sec:nonan}.). Specifically, for these solutions, while the metric tends to a flat metric and the scalar field $\varphi$ tend to zero at infinity, the leading fall-off behaviors of them do not go like $1/r$ (i.e., not of the standard Gaussian fall-off). Also, the scalar field decays and oscillates to zero at infinity. These solutions are presented in Appendix \ref{sec:nonan} for completeness.

\begin{figure}[!htb]\centering
\includegraphics[width=.299\textwidth]{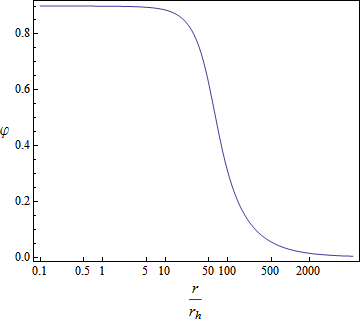}
\includegraphics[width=.305\textwidth]{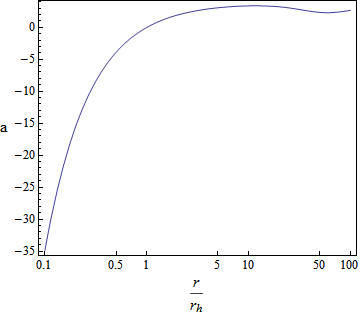}\\
\includegraphics[width=.305\textwidth]{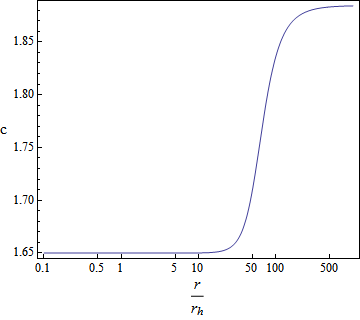}
\includegraphics[width=.33\textwidth]{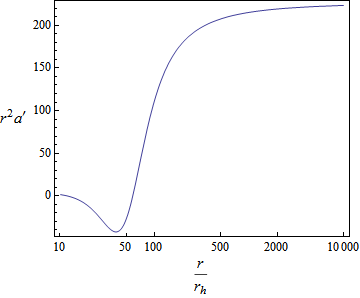}
\caption{Asymptotically flat black hole. $\varphi$ is the mass-varying scalar field, $a$ is the ${}_{tt}$ component of the metric and  $c$ is defined in Eq.~(\ref{cdef}). The mass-varying potential is chosen as $V= m^2 \varphi^4$ with $(m r_h)^2=0.01$. The theory parameters are: $\alpha=1+\alpha_3=1.1,~\beta=\alpha_3+\alpha_4=0.605$. The $k_{3+} =(+\sqrt{\alpha ^2-\beta }-\alpha )/\beta$ branch is chosen for this particular solution. The near horizon ``initial conditions'' are chosen as $c(r_h)=1.65$ and  $\varphi(r_h)=0.9$. $r^2 a'$ becomes constant at large $r$, indicating the metric has the Gaussian fall-off behavior ${1}/{r}$.  }
\label{s2}
\end{figure}

\begin{figure}[!htb]\centering
\includegraphics[width=.339\textwidth]{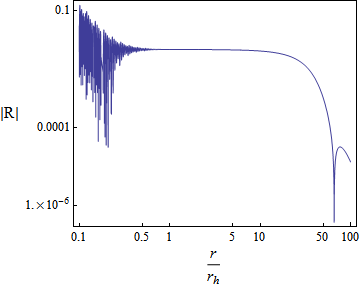}
\includegraphics[width=.36\textwidth]{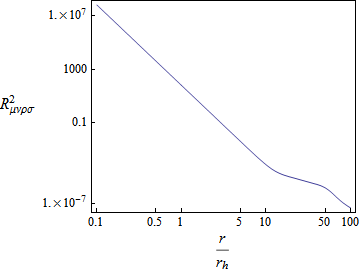}
\caption{$|R|$ and $R_{\mu\nu\kappa\lambda}R^{\mu\nu\kappa\lambda}$ for the two branches of the asymptotically flat black hole solutions detailed in Fig \ref{s2}.  $|R|$ is the absolute vale of Ricci scalar $R$. While the dense spikes on left of the $|R|$ plot are due to numerics, the isolated spike on the right simply indicates $R$ changes the sign at those radii.}
\label{s2-r}
\end{figure}

To get a $1/r$ fall-off behavior for the metric components and the scalar field, we can apply a shooting procedure to tune $c(r_h)$ (or $\varphi(r_h)$) such that $U(\mc{K})\to0$ as $r$ approaches spatial infinity. See Fig.~\ref{s2} for the independent components of a typical solution and see Fig.~\ref{s2-r} for the corresponding Ricci and Kretschmann scalars.

Note that for these solutions $\ud s^2$ and $\ud s_{\eta}^2$ do not necessarily take the standard Minkowski form, ${\rm diag}(-1,1,1,1)$, simultaneously at infinity. They are both asymptotically flat but may differ by a constant scaling of $t$ and $r$.
Nonetheless, being both flat asymptotically, the symmetries of the two geometries are identical at spatial infinity.
If desired, one can obtain a solution where $\ud s^2|_{r\to \infty}\propto \ud s_{\eta}^2|_{r\to \infty}$ by tuning some parameter in the model. See Appendix \ref{sec:sm} for such a solution.

\begin{figure}\centering
\includegraphics[width=.32\textwidth]{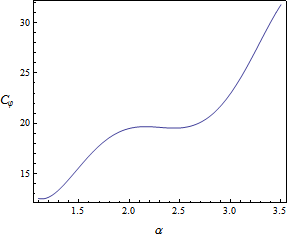}
\includegraphics[width=.325\textwidth]{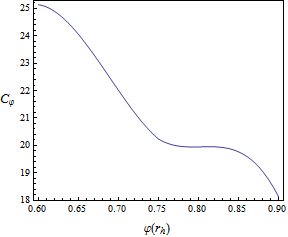}\\
\includegraphics[width=.33\textwidth]{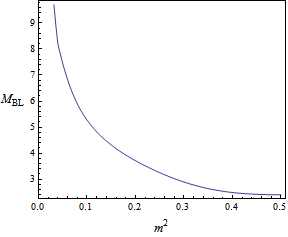}
\includegraphics[width=.325\textwidth]{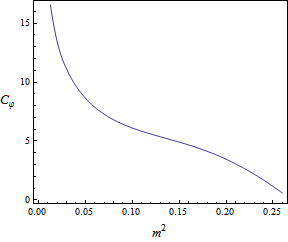}
\caption{The dependence of scalar charge $C_\varphi$ and black hole mass $M_{\rm BL}$ on model parameters. We choose $V=m^2\varphi^4$ with $\bi=0.5\ai^2$. $c(r_h)$ is adjusted such that $U(\mc{K})$ vanishes at infinity to have Gaussian fall-offs for the metric and the scalar field at spatial infinity. For the top-left subfigure we choose $\varphi(r_h)=0.5$ and for the top-right subfigure $\ai=1.1$, with $(m r_h)^2=0.01$ for both of them. The bottom subfigures are the dependence of the ADM mass and the scalar charge on parameter $(mr_h)^2$, with $\ai=1.1$ and $\varphi(r_h)=0.9$.}
\label{scharge}
\end{figure}

The scalar charge $C_\varphi$ is defined at large $r$ as
\be
\varphi(r) = \varphi(\infty) + \frac{C_\varphi}{r} + \mc{O}\left(\frac{1}{r^2}\right)  .
\ee
where $\varphi(\infty)=0$ for our choice of $V(\varphi)$. To extract the black hole mass, we first scale the coordinates $t$ and $r$ to $\tld{t}$ and $\tld{r}$ such that the dynamic metric has the standard Minkowski form at infinity in $\tld{t}$ and $\tld{r}$. Then the black hole (ADM/Komar) mass $M_{BL}$ is defined at large $\tld{r}$ such that the ${}_{tt}$ component of the metric deviates from Minkowski space by
\be
 \delta \tld{g}_{tt} =  \frac{2M_{\rm BL}}{\tld{r}}  + \mc{O}\left(\frac{1}{\tld{r}^2}\right)  
\ee
See Fig \ref{scharge} for how the black hole mass and the scalar charge vary by tuning the graviton mass parameter $m$ and other parameters.

\subsubsection{Asymptotically anti-de Sitter black holes}

As discussed above, to get asymptotically anti-de Sitter black holes, we need to choose $V(\varphi)$ and the theory parameters such that  the $V k_3 \left(\alpha  k_3+2\right)/2$ approaches a positive value at spatial infinity. This can be easily achieved by simply choosing $V=v_0+m^2\varphi^4$, where $v_0$ is a positive constant. A generic choice of $c(r_h)$ and $\varphi(r_h)$ leads to asymptotically anti-de Sitter black holes. We can find sets of $c(r_h)$, $\varphi(r_h)$, $\alpha$ and $\beta$ where both the $k_{3+}$ and $k_{3-}$ branches are asymptotically anti-de Sitter solutions simultaneously, which is similar to the case in weakly asymptotically flat black holes (see Appendix \ref{sec:nonan}). See Fig.~\ref{adss1} and \ref{adss2}.

\begin{figure}\centering
\includegraphics[width=.302\textwidth]{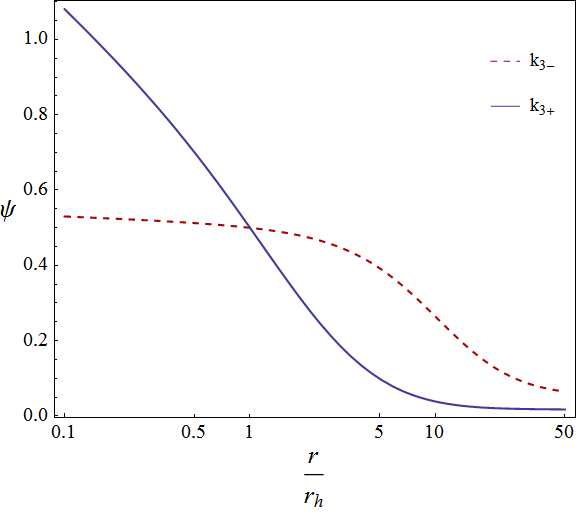}
\includegraphics[width=.3\textwidth]{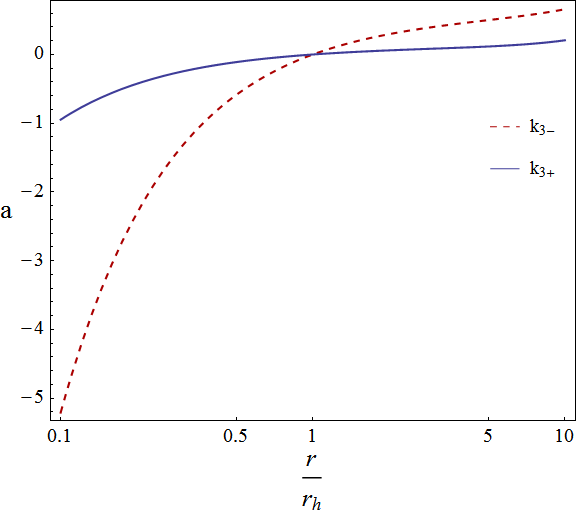}
\includegraphics[width=.308\textwidth]{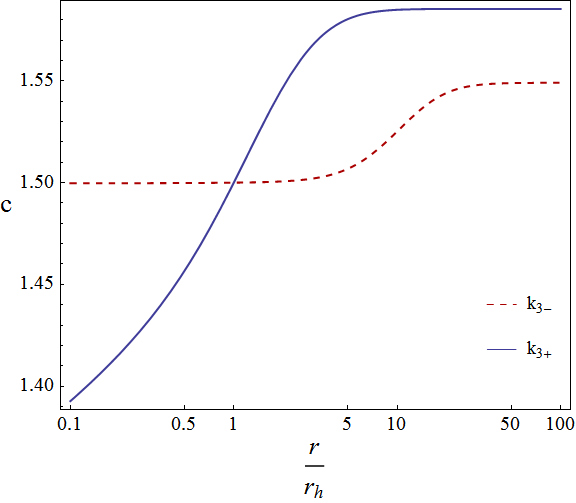}
\caption{Asymptotically anti-de Sitter black holes. $\varphi$ is the mass-varying scalar field, $a$ is the ${}_{tt}$ component of the metric and  $c$ is defined in Eq.~(\ref{cdef}). The mass-varying potential is chosen as $V= v_0+m^2 \varphi^4$ with $v_0 r_h^2=0.01,~(m r_h)^2=0.09$. The theory parameters are: $\alpha=1+\alpha_3=-2.0,~\beta=\alpha_3+\alpha_4=3.6$. There are two branches of solutions: $k_{3\pm} =(\pm\sqrt{\alpha ^2-\beta }-\alpha )/\beta$. The near horizon ``initial conditions'' are chosen as $c(r_h)=1.5, \varphi(r_h)=0.5$ for both branches.}
\label{adss1}
\end{figure}

\begin{figure}\centering
\includegraphics[width=.356\textwidth]{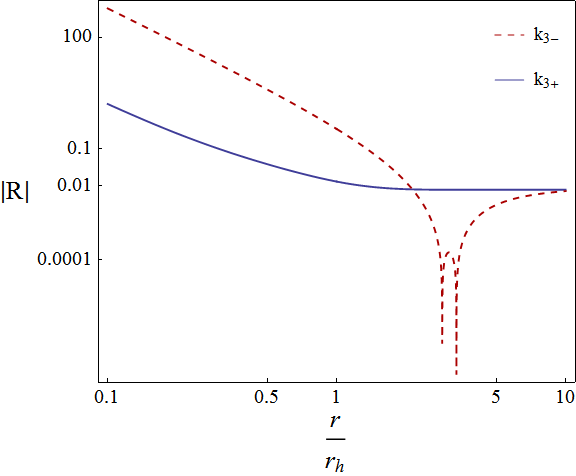}
\includegraphics[width=.38\textwidth]{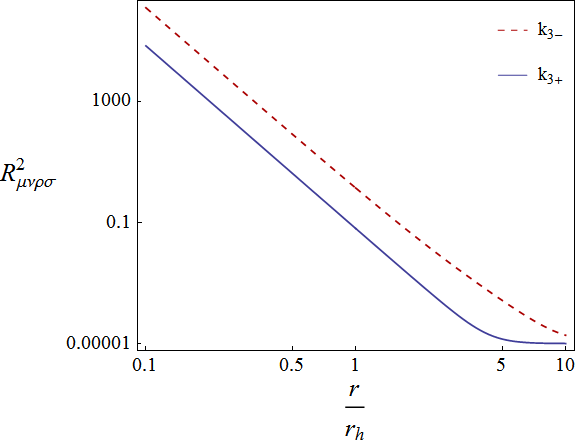}
\caption{$R$ and $R_{\mu\nu\kappa\lambda}R^{\mu\nu\kappa\lambda}$ for the asymptotically anti-de Sitter black hole detailed in Fig \ref{adss1}.}
\label{adss2}
\end{figure}

\section{Conclusions}

For a gravitational theory to be phenomenologically viable, there should be valid black hole solutions, which act as the endpoint of the gravitational collapse of massive stars or larger structures. Also, black hole solutions are important for a sound theoretical understanding of the gravitational theory.

In this paper, we have investigated the static, spherically symmetric black hole solutions in scalar extended massive gravity. We have focused on the simplest example of this class of models, that is, models obtained by promoting the the graviton mass to depending on an extra scalar and adding a canonical kinetic term for this scalar. Similar to the bi-gravity extension of ghost-free massive gravity, we have numerically obtained asymptotically AdS hairy back holes, but not asymptotically de Sitter one. By a hairy, or non-GR, black hole, we mean a black hole geometry that is not Schwarzschild, Schwarzschild-dS or Schwarzschild-AdS. There are two branches of solutions, labeled by $k_{3+}$ and $k_{3-}$ respectively, arising from solving the ${}^r_t$ component of the Einstein equation, which is a quadratic algebraic equation. For a given set of horizon ``initial conditions'' and theory parameters, the two branches can co-exist for weakly asymptotically flat or asymptotically AdS black holes.

We have also obtained asymptotically flat hairy black holes for generic theory parameters. This is in contrast to the bi-gravity extension of ghost-free massive gravity where asymptotically flat hairy black holes only exist for a fine-tuned subset of theory parameters. We have obtained hairy black hole solutions for both weakly flat asymptotics and Gaussian fall-off flat asymptotics. Our focus in the main body of the paper has been on black holes with Gaussian fall-off asymptotics, since for this case one can define a Gaussian flux that can be measured at spatial infinity. We have also shown how the black hole mass and the scalar charge change with model parameters.

We have focused on the simplest model where $W(\varphi)=0$ and $V(\varphi)$ is a monomial (or monomial plus a constant), which already has a rich black hole solution space. Not presented in this paper, we have also investigated cases where $W(\varphi)\neq 0$ and/or $V(\varphi)$ is a polynomial. For the limited such cases we have considered, it does not seem to give rise to dramatically new features. Of course, one may also promote $\alpha_3$ and $\alpha_4$ to depending on the scalar field, or consider models such as generalized quasi-dilation \cite{DeFelice:2013dua}, which is likely to give rise to some new features, as the ${}^r_t$ component of the Einstein equation for these models are dramatically different from the counterpart here (Eq.~(\ref{fenlei})). We leave this for future work.

If some added fields can generate some new branches of (hairy) black holes in a gravitational theory, it might be expected that the GR-like branch (if there is) may be unstable, since the added fields with nontrivial configurations inject gravitating energy into the system. For our case, it seems to be more likely the case, as the GR-like solutions are already unstable in the limit of ghost-free massive gravity. However, we shall leave the stability analysis for future work.

\vskip 20pt

\textbf{Acknowledgments}
We would like to thank Yun-Song Piao and Thomas Sotiriou for discussions. AJT and SYZ acknowledges support from DOE grant DE-SC0010600. DJW is supported by NSFC, No.~11222546, and National Basic Research Program of China, No.~2010CB832804, and the Strategic Priority Research Program of Chinese Academy of Sciences, No.~XDA04000000. SYZ would also like to thank Kavli Institute for Theoretical Physics China at the Chinese Academy of Sciences for hospitality during part of this work.

\appendix

\section*{APPENDICES}

\subsection{Nonzero components of the Einstein equation}
\label{GTXlist}

Here we list all the nonzero components of $G^\mu_\nu$, $T^\mu_\nu$ and $X^\mu_\nu$ (indices being raised by $g^{\mu\nu}$) for ansatz (\ref{metric})--(\ref{varphian}):
\bea
 G_t^t  &=& \frac{{  2cfa'f' + 4acff'' - acf'^2  - 2afc'f' - 4c^2 f}}{{4c^2 f^2 }} \\
 G_r^t  &=& \frac{{ bf {c'f' - 2b c ff''}  + b cf'^2 }}{{2c^2 f^2 }} \\
 G_r^r  &=& \frac{{ {2fa'f' + af'^2}  - 4cf}}{{4cf^2 }} \\
 G_\theta ^\theta   &=& G_\phi ^\phi   = \frac{{ { {2cfa'f' - afc'f'}  + 2acff''} +   2cf^2a'' - f^2a'c' - acf'^2 }}{{4c^2 f^2 }}
\\
T^{t}_{t}&=&T^{\theta}_{\theta}=T^{\varphi}_{\varphi}=
- \frac{{a\varphi '^2 }}{{2c}} - W\\
T^{r}_{r}&=&\frac{{a\varphi '^2 }}{{2c}} - W\\
T^{t}_{r}&=&{\frac{{b\varphi '^2 }}{{c}}},
\\
X^{t}_{t}&=&\frac{{\left(b^2 n^2 k_1-k_2\right)\left( {\beta {{k_3}}^2  + 2\alpha {{k_3}} + 1} \right)}}{{2(b^2 n^2  - 1)}}+ \frac{ {k_3} (\alpha  {k_3}+2)}{2},\\
X^{r}_{r}&=&\frac{{\left(b^2 n^2 k_2-k_1\right)\left( {\beta {{k_3}}^2  + 2\alpha {{k_3}} + 1} \right)}}{{2(b^2 n^2  - 1)}}+\frac{ {k_3} (\alpha  {k_3}+2)}{2},\\
X^{r}_{t}&=&-X^{t}_{r}=-\frac{b }{2c}\left(\beta  {k_3}^2+2 \alpha  {k_3}+1\right),\\
X^{\theta}_{\theta}&=&X^{\varphi}_{\varphi}=\frac{\alpha (k_1{{k_2}} + k_1{{k_3}}+ k_2k_3) + \beta k_1{{k_2k_3}}   +k_1+ {{k_2}} + {{k_3}}}{2}
\eea

\subsection{Weakly asymptotically flat black holes}
\label{sec:nonan}

\begin{figure}[!htb]\centering
\includegraphics[width=.3\textwidth]{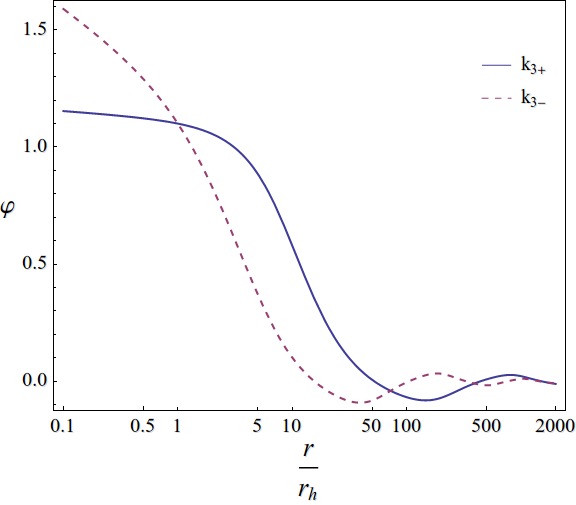}
\includegraphics[width=.308\textwidth]{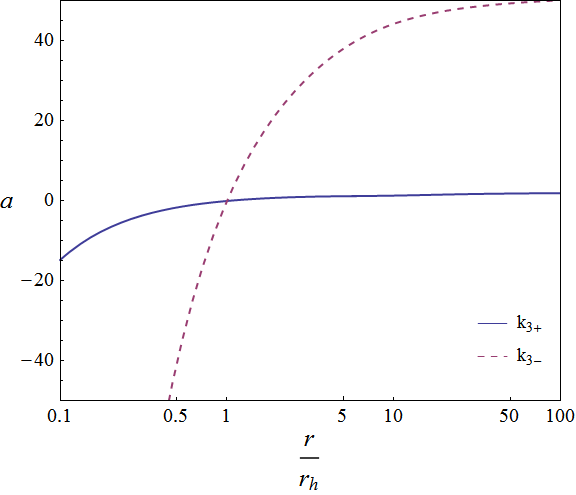}\\
\includegraphics[width=.305\textwidth]{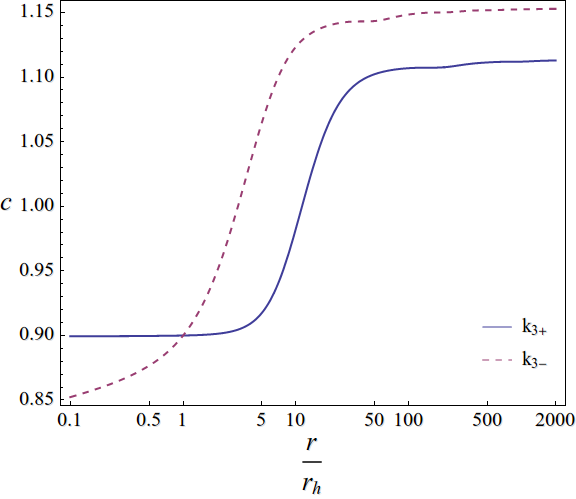}
\includegraphics[width=.336\textwidth]{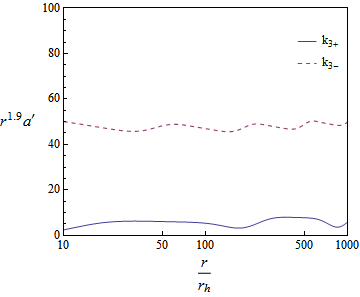}
\caption{Weakly asymptotically flat black hole. $\varphi$ is the mass-varying scalar field, $a$ is the ${}_{tt}$ component of the metric and  $c$ is defined in Eq.~(\ref{cdef}). The mass-varying potential is chosen as $V= m^2 \varphi^4$ with $(m r_h)^2=0.09$. The theory parameters are: $\alpha=1+\alpha_3=1.5,~\beta=\alpha_3+\alpha_4=0.5$. There are two branches of solutions: $k_{3\pm} =(\pm\sqrt{\alpha ^2-\beta }-\alpha )/\beta$. The near horizon ``initial conditions'' are chosen as $c(r_h)=0.9, \varphi(r_h)=1.1$ for both branches. }
\label{fs1}
\end{figure}

\begin{figure}[!htb] \centering
\includegraphics[width=.34\textwidth]{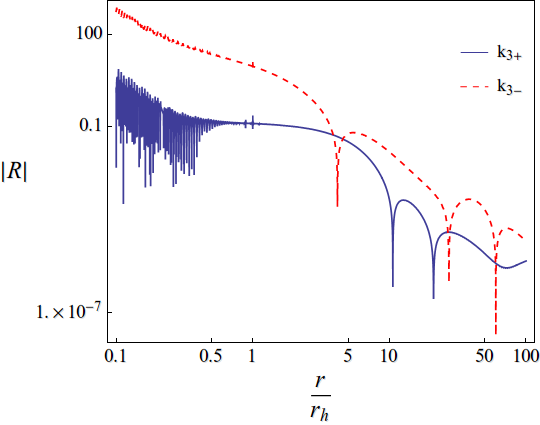}
\includegraphics[width=.36\textwidth]{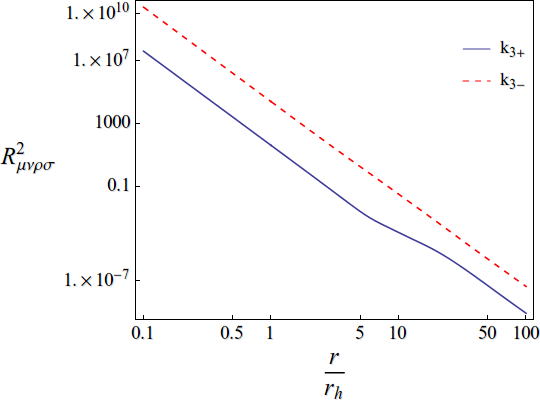}
\caption{$|R|$ and $R_{\mu\nu\kappa\lambda}R^{\mu\nu\kappa\lambda}$ for the two branches of the weakly asymptotically flat black hole solutions detailed in Fig \ref{fs1}.  $|R|$ is the absolute vale of Ricci scalar $R$. While the dense spikes on left of the $|R|$ plot are due to numerics, the isolated spikes on the right simply indicates $R$ changes the sign at those radii.}
\label{fs2}
\end{figure}

There are several versions of definition for asymptotical flatness. For our purposes, we make use of a set of Cartesian coordinates $(t,x,y,z)$. Then, an asymptotically flat spacetime simply requires that the metric $g_{\mu\nu}$ falls off like
\be
\lim_{r\to \infty}g_{\mu\nu} = {\rm const} + \mc{O}\left(r^{-p}\right)
,~~
\lim_{r\to \infty}\pd_\ri g_{\mu\nu} =  \mc{O}\left( r^{-p-1}\right)
,~~
\lim_{r\to \infty}\pd_\ri\pd_\si g_{\mu\nu} =  \mc{O}\left( r^{-p-2}\right)
\ee
where $r=\sqrt{x^2+y^2+z^2}$. When there are additional fields, a similar requirement is imposed for those additional fields. A weaker version requires that $p>1/2$. This is because the gravitational energy diverges when $p\leq 1/2$, as at large $r$ the gravitational energy goes like $\int r^2\ud r (\pd g)^2$. A stronger version requires $p=1$. When there is a $1/r$ term in the metric (and additional fields), we have the standard Gaussian flux, in which case the coefficient of the $1/r$ term is a charge that can be observed at spatial infinity.

\begin{figure}\centering
\includegraphics[width=.40\textwidth]{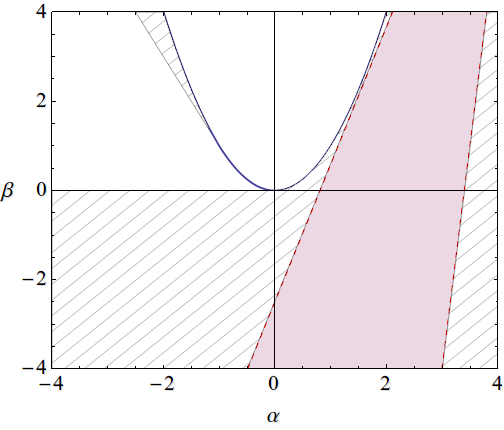}
\caption{Parameter space of $\alpha$ and $\beta$ that lead to weakly asymptotically flat hairy black holes (shaded light red area) for the $k_{3+}$ branch with $c(r_h)=0.9$, $\varphi(r_h)=1.1$). $V=m^2\varphi^4$ with $(m r_h)^2=0.09$. The white region is excluded by the reality condition, i.e., Eq.~(\ref{abpara}) with the plus sign, and the slashed area are excluded by laboriously running numerical integrations for different parameters. The parameter space of the $k_{3-}$ branch seems to be only constrained by the reality condition Eq.~(\ref{abpara}) with the minus sign.}
\label{abarea}
\end{figure}

As mentioned in Section \ref{sec:afs}, for the black hole with a generic choice of $c(r_h)$ and $\varphi({r_h})$, the leading fall-offs of the metric and $\varphi$ are $p<1$. But they still satisfy the weak requirement of $p>1/2$, so a generic black hole for this case is asymptotically flat in the weak sense. See Fig.~\ref{fs1} and \ref{fs2}. For a specific choice of $c(r_h)$ and $\varphi({r_h})$, it seems that there exists a black hole solution for almost any $\ai$ and $\bi$ allowed by the reality condition of $k_3$ (i.e., Eq.~(\ref{rck3})) for one branch of $k_3$. For the other branch, there are usually further restrictions on $\ai$ and $\bi$ to generate a black hole solution. See Fig.~\ref{abarea} for an example.

\subsection{Asymptotically flat black hole: simultaneously Minkowski}
\label{sec:sm}

\begin{figure}\centering
\includegraphics[width=.305\textwidth]{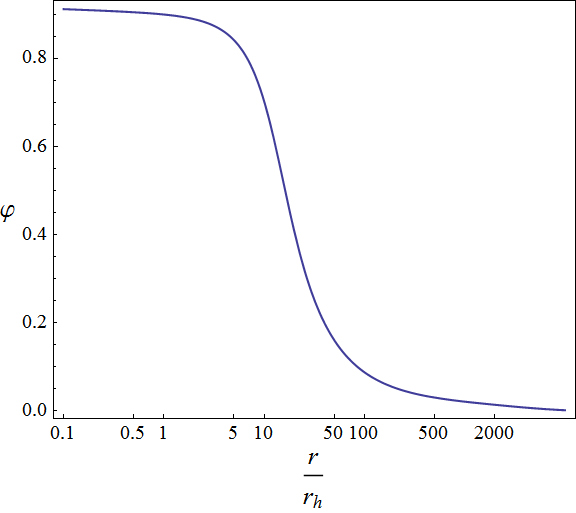}
\includegraphics[width=.308\textwidth]{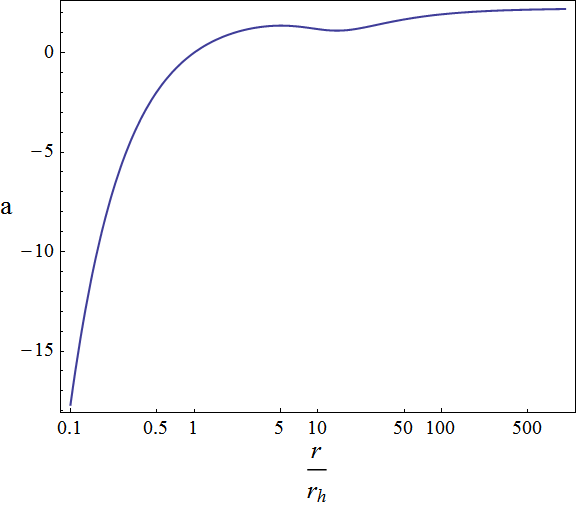}
\includegraphics[width=.305\textwidth]{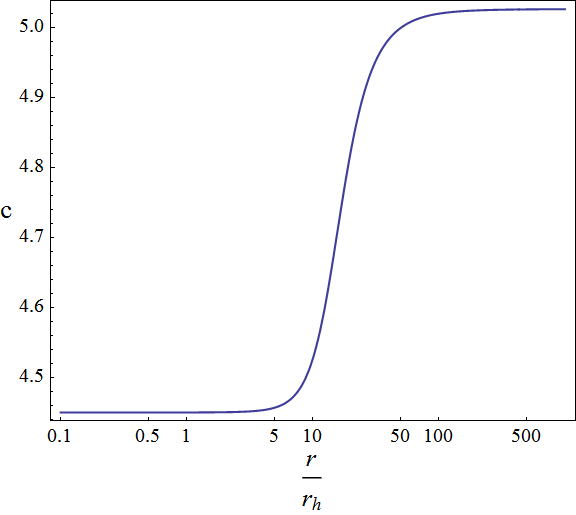}
\caption{Asymptotically flat black hole (in the stronger sense) that is proportional to the Minkowski form at spatial infinity. We choose the $k_{3+} =(+\sqrt{\alpha ^2-\beta }-\alpha )/\beta$ branch. $\varphi$ is the mass-varying scalar field, $a$ is the ${}_{tt}$ component of the metric and  $c$ is defined in Eq.~(\ref{cdef}). The mass-varying potential is chosen as $V= m^2 \varphi^4$ with $(m r_h)^2=0.01$. The theory parameters are: $\alpha=1+\alpha_3=-2,~\beta=\alpha_3+\alpha_4=2.04$. The near horizon ``initial conditions'' are chosen as $c(r_h)=4.45, \varphi(r_h)=0.9$ for both branches. }
\label{n-s1}
\end{figure}

\begin{figure}\centering
\includegraphics[width=.35\textwidth]{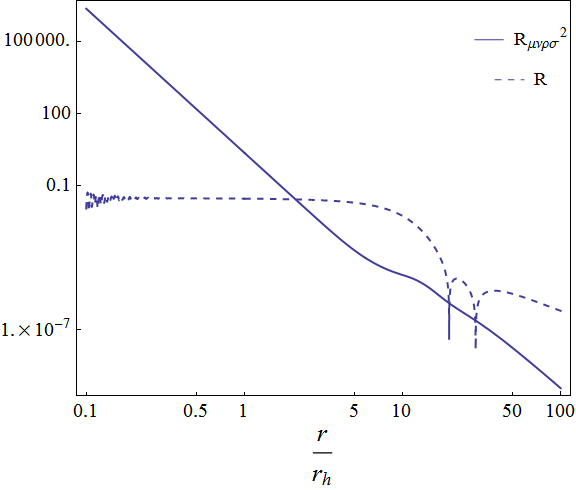}
\includegraphics[width=.324\textwidth]{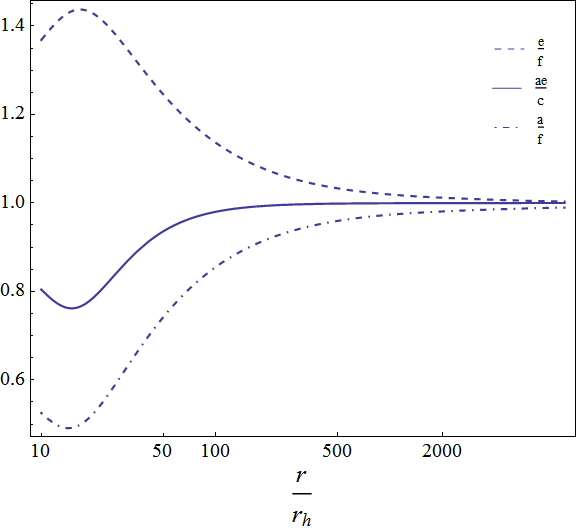}
\caption{The left subfigure is the Ricci scalar $|R|$ and Kretschmann scalar $R_{\mu\nu\kappa\lambda}R^{\mu\nu\kappa\lambda}$ for the the asymptotically flat black hole solutions detailed in Fig \ref{n-s1}.  $|R|$ is the absolute vale of Ricci scalar $R$. While the dense spikes on left of the $|R|$ plot are due to numerics, the isolated spikes on the right simply indicates $R$ changes the sign at those radii. In the right subfigure, we plot the metric components (see Eq.~(\ref{metric})) for the solution in Fig.~\ref{n-s1}.}
\label{n-s1-r}
\end{figure}

For asymptotically flat black holes (in the stronger sense), the dynamical metric $\ud s^2$ may not be of the standard Minkowski form (i.e. the form of the reference metric $\ud s_\eta^2$) at infinity. By tuning another parameter in the model, one can make them proportional to each other at infinity. For example,  given an asymptotically flat solution, one can further tune $\ai$ to achieve this.  An example is given in Fig.~\ref{n-s1} and \ref{n-s1-r}.  To see $\ud s^2$ is proportional to the standard Minkowski form at spatial infinity, we plot the ${}_{tt}$ component ($a$) and ${}_{rr}$ component ($e$) of the metric against the ${}_{\theta\theta}$ component ($f$), which should be equal to 1. In addition, we also plot ${ae}/{c}$ at large $r$, which should also be equal to $1$ if $b|_{r=\infty}=0$. See Fig.~\ref{n-s1-r}.

\begin{thebibliography}{99}
\bibitem{deRham:2014zqa}
  C.~de Rham,
  Living Rev.\ Rel.\  {\bf 17}, 7 (2014)
  [arXiv:1401.4173 [hep-th]].

\bibitem{Hinterbichler:2011tt}
  K.~Hinterbichler,
  Rev.\ Mod.\ Phys.\  {\bf 84}, 671 (2012)
  [arXiv:1105.3735 [hep-th]].

\bibitem{Riess:1998cb}
  A.~G.~Riess {\it et al.}  [Supernova Search Team Collaboration],
  Astron.\ J.\  {\bf 116}, 1009 (1998)
  [astro-ph/9805201].

\bibitem{Perlmutter:1998np}
  S.~Perlmutter {\it et al.}  [Supernova Cosmology Project Collaboration],
  Astrophys.\ J.\  {\bf 517}, 565 (1999)
  [astro-ph/9812133].

\bibitem{Fierz:1939ix}
  M.~Fierz and W.~Pauli,
  Proc.\ Roy.\ Soc.\ Lond.\ A {\bf 173}, 211 (1939).

\bibitem{Boulware:1973my}
  D.~G.~Boulware and S.~Deser,
  Phys.\ Rev.\ D {\bf 6}, 3368 (1972).

\bibitem{deRham:2010ik}
  C.~de Rham, G.~Gabadadze,
  Phys.\ Rev.\  {\bf D82}, 044020 (2010),
  [arXiv:1007.0443].

\bibitem{deRham:2010kj}
  C.~de Rham, G.~Gabadadze and A.~J.~Tolley,
  Phys.\ Rev.\ Lett.\  {\bf 106}, 231101 (2011),
  [arXiv:1011.1232].

\bibitem{Hassan:2011hr}
  S.~F.~Hassan and R.~A.~Rosen,
  Phys.\ Rev.\ Lett.\  {\bf 108}, 041101 (2012)
  [arXiv:1106.3344 [hep-th]].

\bibitem{Hassan:2011ea}
  S.~F.~Hassan and R.~A.~Rosen,
  JHEP {\bf 1204} (2012) 123
  [arXiv:1111.2070 [hep-th]].

\bibitem{Nieuwenhuizen:2011sq}
  T.~M.~Nieuwenhuizen,
  Phys.\ Rev.\ D {\bf 84}, 024038 (2011)
  [arXiv:1103.5912 [gr-qc]].

\bibitem{Koyama:2011xz}
  K.~Koyama, G.~Niz and G.~Tasinato,
  Phys.\ Rev.\ Lett.\  {\bf 107}, 131101 (2011)
  [arXiv:1103.4708 [hep-th]].

\bibitem{Koyama:2011yg}
  K.~Koyama, G.~Niz and G.~Tasinato,
  Phys.\ Rev.\ D {\bf 84}, 064033 (2011)
  [arXiv:1104.2143 [hep-th]].

\bibitem{Gruzinov:2011mm}
  A.~Gruzinov and M.~Mirbabayi,
  Phys.\ Rev.\ D {\bf 84}, 124019 (2011)
  [arXiv:1106.2551 [hep-th]].

\bibitem{Comelli:2011wq}
  D.~Comelli, M.~Crisostomi, F.~Nesti and L.~Pilo,
  Phys.\ Rev.\ D {\bf 85}, 024044 (2012)
  [arXiv:1110.4967 [hep-th]].

\bibitem{Berezhiani:2011mt}
  L.~Berezhiani, G.~Chkareuli, C.~de Rham, G.~Gabadadze and A.~J.~Tolley,
  Phys.\ Rev.\ D {\bf 85}, 044024 (2012)
  [arXiv:1111.3613 [hep-th]].

\bibitem{Volkov:2012wp}
  M.~S.~Volkov,
  Phys.\ Rev.\ D {\bf 85}, 124043 (2012)
  [arXiv:1202.6682 [hep-th]].

\bibitem{Baccetti:2012ge}
  V.~Baccetti, P.~Martin-Moruno and M.~Visser,
  JHEP {\bf 1208}, 108 (2012)
  [arXiv:1206.4720 [gr-qc]].

\bibitem{Cai:2012db}
  Y.~F.~Cai, D.~A.~Easson, C.~Gao and E.~N.~Saridakis,
  Phys.\ Rev.\ D {\bf 87}, 064001 (2013)
  [arXiv:1211.0563 [hep-th]].

\bibitem{Berezhiani:2013dw}
  L.~Berezhiani, G.~Chkareuli and G.~Gabadadze,
  Phys.\ Rev.\ D {\bf 88}, 124020 (2013)
  [arXiv:1302.0549 [hep-th]].

\bibitem{Mirbabayi:2013sva}
  M.~Mirbabayi and A.~Gruzinov,
  Phys.\ Rev.\ D {\bf 88}, 064008 (2013)
  [arXiv:1303.2665 [hep-th]].

\bibitem{Volkov:2013roa}
  M.~S.~Volkov,
  Class.\ Quant.\ Grav.\  {\bf 30}, 184009 (2013)
  [arXiv:1304.0238 [hep-th]].

\bibitem{Tasinato:2013rza}
  G.~Tasinato, K.~Koyama and G.~Niz,
  Class.\ Quant.\ Grav.\  {\bf 30}, 184002 (2013)
  [arXiv:1304.0601 [hep-th]].

\bibitem{Babichev:2013una}
  E.~Babichev and A.~Fabbri,
  Class.\ Quant.\ Grav.\  {\bf 30}, 152001 (2013)
  [arXiv:1304.5992 [gr-qc]].

\bibitem{Brito:2013wya}
  R.~Brito, V.~Cardoso and P.~Pani,
  Phys.\ Rev.\ D {\bf 88}, no. 2, 023514 (2013)
  [arXiv:1304.6725 [gr-qc]].

\bibitem{Berezhiani:2013dca}
  L.~Berezhiani, G.~Chkareuli, C.~de Rham, G.~Gabadadze and A.~J.~Tolley,
  Class.\ Quant.\ Grav.\  {\bf 30}, 184003 (2013)
  [arXiv:1305.0271 [hep-th]].

\bibitem{Brito:2013xaa}
  R.~Brito, V.~Cardoso and P.~Pani,
  Phys.\ Rev.\ D {\bf 88}, 064006 (2013)
  [arXiv:1309.0818 [gr-qc]].

\bibitem{Kodama:2013rea}
  H.~Kodama and I.~Arraut,
  PTEP {\bf 2014}, 023E02 (2014)
  [arXiv:1312.0370 [hep-th]].

\bibitem{Kobayashi:2015yda}
  T.~Kobayashi, M.~Siino, M.~Yamaguchi and D.~Yoshida,
  arXiv:1509.02096 [gr-qc].

\bibitem{Addazi:2014mga}
  A.~Addazi and S.~Capozziello,
  Int.\ J.\ Theor.\ Phys.\  {\bf 54} (2015) 6,  1818
  [arXiv:1407.4840 [gr-qc]].

\bibitem{Babichev:2014oua}
  E.~Babichev and A.~Fabbri,
  Phys.\ Rev.\ D {\bf 89}, no. 8, 081502 (2014)
  [arXiv:1401.6871 [gr-qc]].

\bibitem{Babichev:2014fka}
  E.~Babichev and A.~Fabbri,
  JHEP {\bf 1407}, 016 (2014)
  [arXiv:1405.0581 [gr-qc]].

\bibitem{Volkov:2014ooa}
  M.~S.~Volkov,
  Lect.\ Notes Phys.\  {\bf 892}, 161 (2015)
  [arXiv:1405.1742 [hep-th]].

\bibitem{Babichev:2014tfa}
  E.~Babichev and A.~Fabbri,
  Phys.\ Rev.\ D {\bf 90}, 084019 (2014)
  [arXiv:1406.6096 [gr-qc]].

\bibitem{Babichev:2015xha}
  E.~Babichev and R.~Brito,
  arXiv:1503.07529 [gr-qc].

\bibitem{Enander:2015}
J.~Enander and E.~Mortsell,
arXiv:1507.00912 [astro-ph.CO].

\bibitem{Deffayet:2011rh}
  C.~Deffayet and T.~Jacobson,
  Class.\ Quant.\ Grav.\  {\bf 29}, 065009 (2012)
  [arXiv:1107.4978 [gr-qc]].

\bibitem{Hassan:2011zd}
  S.~F.~Hassan and R.~A.~Rosen,
  JHEP {\bf 1202}, 126 (2012)
  [arXiv:1109.3515 [hep-th]].

\bibitem{Bekenstein:1996pn}
  J.~D.~Bekenstein,
  In *Moscow 1996, 2nd International A.D. Sakharov Conference on physics* 216-219
  [gr-qc/9605059].

\bibitem{Herdeiro:2015waa}
  C.~A.~R.~Herdeiro and E.~Radu,
  Int.\ J.\ Mod.\ Phys.\ D {\bf 24}, no. 09, 1542014 (2015)
  [arXiv:1504.08209 [gr-qc]].

\bibitem{D'Amico:2012zv}
  G.~D'Amico, G.~Gabadadze, L.~Hui and D.~Pirtskhalava,
  Phys.\ Rev.\ D {\bf 87}, 064037 (2013)
  [arXiv:1206.4253 [hep-th]].

\bibitem{D'Amico:2011jj}
  G.~D'Amico, C.~de Rham, S.~Dubovsky, G.~Gabadadze, D.~Pirtskhalava and A.~J.~Tolley,
  Phys.\ Rev.\ D {\bf 84}, 124046 (2011)
  [arXiv:1108.5231 [hep-th]].

\bibitem{Huang:2012pe}
  Q.~G.~Huang, Y.~S.~Piao and S.~Y.~Zhou,
  Phys.\ Rev.\ D {\bf 86}, 124014 (2012)
  [arXiv:1206.5678 [hep-th]].

\bibitem{Huang:2013mha}
  Q.~G.~Huang, K.~C.~Zhang and S.~Y.~Zhou,
  JCAP {\bf 1308}, 050 (2013)
  [arXiv:1306.4740 [hep-th]].

\bibitem{DeFelice:2013dua}
  A.~De Felice, A.~Emir G\"{u}mr\"{u}k\c{c}\"{u}o\u{g}lu and S.~Mukohyama,
  Phys.\ Rev.\ D {\bf 88}, no. 12, 124006 (2013)
  [arXiv:1309.3162 [hep-th]].

\bibitem{Andrews:2013ora}
  M.~Andrews, G.~Goon, K.~Hinterbichler, J.~Stokes and M.~Trodden,
  Phys.\ Rev.\ Lett.\  {\bf 111}, no. 6, 061107 (2013)
  [arXiv:1303.1177 [hep-th]].

\bibitem{Mukohyama:2014rca}
  S.~Mukohyama,
  JCAP {\bf 1412}, no. 12, 011 (2014)
  [arXiv:1410.1996 [hep-th]].








\end{thebibliography}
\end{document}